\documentclass[preprint,12]{mn2e}

\usepackage{epsfig}
\usepackage{color}

\def\lsim{\mathrel{\rlap{\lower 4pt \hbox{\hskip 1pt $\sim$}}\raise 1pt\hbox {$<$}}}
\def\gsim{\mathrel{\rlap{\lower 4pt \hbox{\hskip 1pt $\sim$}}\raise 1pt\hbox {$>$}}}
\def\Msun{\mbox{\,M$_\odot$}}

\title[Explosive Common-Envelope Ejection]
{Explosive Common-Envelope Ejection: Implications for Gamma-Ray
Bursts and Low-Mass Black-Hole Binaries}

\author[Podsiadlowski et al.]  {Philipp
  Podsiadlowski$^{1}$\thanks{E-mail: podsi@astro.ox.ac.uk}, Natasha
  Ivanova$^{2,3}$, Stephen Justham$^{1,4}$, Saul
  Rappaport$^{5}$\\ 
$^{1}$ {\it Dept.\ of Astronomy, Oxford University,
    Oxford, OX1 3RH, UK}\\ 
$^{2}$ {\it CITA, University of Toronto, 60
    St.\ George, Toronto, ON M5S 3H8, Canada}\\ 
$^{3}$ {\it Department of Physics, University of Alberta, Edmonton, AB T6G 2G7, Canada}\\
$^{4}$ {\it Kavli Institute for Astronomy and Astrophysics, Peking University,
    Beijing 100871, China}\\ 
$^{5}$ {\it Department of Physics and Kavli Institute for Astrophysical
    Research, Massachusetts Institute of Technology,
    Cambridge, MA 02139} 
} 
\date{\today}

\volume{000}

\pagerange{1\,--\,8} \pubyear{2010}

\begin{document}

\label{firstpage}

\maketitle
\begin{abstract}
We present a new mechanism for the ejection of a common envelope in a
massive binary, where the energy source is nuclear energy rather than
orbital energy. This can occur during the slow merger of a massive
primary with a secondary of $1-3\,$M$_\odot$ when the primary has
already completed helium core burning. We show that, in the final
merging phase, hydrogen-rich material from the secondary can be
injected into the helium-burning shell of the primary. This leads to a
nuclear runaway and the explosive ejection of both the hydrogen and
the helium layer, producing a close binary containing a CO star and a
low-mass companion. We argue that this presents a viable scenario to
produce short-period black-hole binaries and long-duration gamma-ray
bursts (LGRBs). We estimate a LGRB rate of $\sim 10^{-6}$\,yr$^{-1}$
at solar metallicity, which implies that this may account for a
significant fraction of all LGRBs, and that this rate should be higher
at lower metallicity.

\end{abstract}



\section{Introduction}

Roughly half of the known short-period black-hole binaries have
low-mass companions (Lee, Brown \& Wijers 2002). However, it has been
realized for more than a decade now that such systems are difficult to
form (see the discussion in Podsiadlowski, Rappaport \& Han 2003 [PRH]
and further references therein). The problem is that, in order to
produce a short-period system (i.e., $\la 15$ hrs) with a low-mass
donor star, the progenitor system has to pass through a
common-envelope (CE) phase where the binary's period is reduced from a
typical period of several years to less than a few days. In the
standard model for CE evolution (Paczy\'nski 1976), this requires that
the orbital energy released in the spiral-in process is sufficient to
eject the massive envelope of the primary. For a low-mass companion
(i.e., $\la 2~M_\odot$), this is energetically difficult, if not
impossible, in particular considering the large binding energies of
the massive envelope of the primaries (Dewi \& Tauris 2001;
PRH).\footnote{In a recent study by Yungelson et al.\ (2006), as well
  as in some other studies, this problem did not seem to
  arise. However, these authors used a prescription for the binding
  energy of the envelope that {\em underestimates} the binding energy
  by up to a factor of 10 compared to the actual binding energies
  calculated from realistic models for the structure of massive
  supergiants (Dewi \& Tauris 2001; PRH).}

This problem has led to the suggestion of a number of more exotic
formation scenarios for low-mass black-hole binaries, e.g., involving a
triple scenario (Eggleton \& Verbunt 1986) or the formation of the
companion from a disrupted envelope (Podsiadlowski, Cannon \& Rees
1995; PRH). Alternatively, the low-mass
black-hole binaries could descend from intermediate-mass systems
(Justham, Rappaport \& Podsiadlowski 2006), as is the case for the
majority of low-mass neutron-star binaries (Pfahl, Rappaport \&
Podsiadlowski 2003). In this paper, we present a new mechanism for the
ejection of the common envelope: ``explosive common-envelope
ejection'', involving nuclear rather than orbital energy, which can be
highly efficient in ejecting a massive envelope even if the companion is a
relatively low-mass star.

A particularly interesting black-hole binary is the well studied
system GRO J1655--40 (Nova Scorpii 1994). It has an orbital period of
2.6\,d and a black hole with a mass $\sim 5.4\,M_\odot$ (Beer \&
Podsiadlowski 2002). Israelian et al.\ (1999) claimed that the
secondary in this system has been highly enriched with the products of
explosive nucleosynthesis (e.g., Mg, Si, S, Ti) produced in the
supernova explosion that produced the black hole.\footnote{In this
  context, we refer the reader to two related studies: one by Foellmi,
  Dall \& Depagne (2007) challenging the original claim of these
  overabundances, and one by Gonz\'alez Hern\'andez, Rebolo \&
  Israelian (2008) re-affirming them.}  Based on the actual abundance
ratios, Podsiadlowski et al.\ (2002) found some tentative evidence that these
are better explained by an energetic supernova explosion, a hypernova,
with a typical ejecta energy of $\ga 10^{52}$\,ergs (i.e., 10 times
the energy in a `typical' supernova). This may suggest that the
formation of the black hole in this system could have been accompanied
by a long-duration gamma-ray burst (LGRB).  However, in this case it
is puzzling why the black-hole progenitor would have been rapidly
rotating as required in the collapsar model for LGRBs (Woosley 1993;
MacFadyen \& Woosley 1999) since the core of the primary should have
been spun down rather than have been spun up during its evolution
(Heger, Woosley \& Spruit 2005).  As we will show later in this paper,
in the case of explosive CE ejection, this problem does not arise,
possibly linking the formation of compact black-hole binaries having
low-mass secondaries to LGRBs.\footnote{Brown et al.\ (2000) were the
  first to suggest that the formation of the black hole in GRO
  J1655-40 was associated with an LGRB and proposed a general link
  between low-mass black-hole binaries and LGRBs (see, in particular,
  Brown, Lee \& Moreno M\'endez 2007). Similar to the present study,
  they suggest late Case C mass transfer for the progenitor systems,
  but argue for the spin-up of the cores rather than spin-down during
  the CE phase, invoking tidal locking, an assumption that remains to
  be proven.}

This process of explosive CE was discovered in a systematic study of
the slow merger of massive stars (Ivanova 2002; Ivanova \&
Podsiadlowski 2003; Ivanova \& Podsiadlowski 2010), where it was found
that, in some cases in the late stage of the spiral-in process,
hydrogen-rich material could be mixed into the helium-burning shell
leading to a thermonuclear runaway which released enough energy to
eject both the hydrogen and the helium envelope (see Figure~1 for a
schematic representation), possibly explaining why to date all
supernovae associated with LGRBs appear to be Type Ic supernovae
(i.e., supernovae without hydrogen and helium in their spectrum;
cf. Podsiadlowski et al.\ 2004).\footnote{We note that Siess \& Livio
(1999a,b) were the first to point out the potential importance of
nuclear energy on the CE ejection process in their modelling of
the dissolution of planets/brown dwarfs in red-giant stars.}

In this paper, we will first discuss the numerical method in Section~2,
and the physics of the process of explosive CE ejection in
Section~3. In Section~4 we apply it to the formation of low-mass
black-hole binaries and LGRBs, and end with a broader discussion in
Section~5.

\section{The Numerical Code}

To model the merging of the two stars, we used a modified
one-dimensional Henyey-type stellar evolution code (based on
Kippenhahn, Weigert \& Hofmeister 1967), which has recently been
updated (Podsiadlowski, Rappaport \& Pfahl 2002). It uses OPAL
opacities (Rogers \& Iglesias 1992), supplemented by molecular
opacities from Alexander \& Ferguson (1994) at low temperatures and a
mixing-lenth parameter $\alpha=2$. Following the calibration by
Schr\"oder, Pols \& Eggleton (1997) and Pols et al.\ (1997), we
generally assume 0.25 pressure scale heights of convective
overshooting, unless stated otherwise.

In the calculation of the spiral-in and merger phase we followed the
angular-momentum transport in the envelope of the primary, using
prescriptions similar to those given by Heger \& Langer (1998), where
the source of the angular momentum is the orbital decay of the
immersed binary. This is important for determining the averaged
distribution of the heating sources inside the common envelope. The
envelope heating is provided by the dissipation of the kinetic energy
of the differentially rotating shells due to viscous friction.  Note
that this energy is not necessarily released in the same place where
the angular momentum has been deposited initially.  For modelling the
latter, we included a prescription that simulates the frictional
energy input during the spiral-in phase of the companion star similar
to the prescription given by Meyer \& Meyer-Hofmeister (1979) (also
see Podsiadlowski 2001).

Once the immersed secondary fills its own critical potential surface
(i.e., its equivalent ``Roche lobe'' defined by the effective
potential of the core-secondary pair) within the common envelope, we
simulate the mass transfer guided by the detailed stream--core
simulations presented in Ivanova, Podsiadlowski \& Spruit (2002).  We
use their equation (28) to determine the depth to which the stream
penetrates the hydrogen-exhausted core of the primary and vary the
parameter $k$ to take into account uncertainties in the modelling of
the stream--core impact.\footnote{ 
The parameter $k$ defines the stream penetration efficiency and
depends on the amount of entropy that is generated in the stream--core
impact: a larger $k$ corresponds to the case where the ambient medium
near the core has a larger pressure and temperature gradient resulting
in the generation of stronger shocks and faster dissipation of the
stream.}

In each time step, we determine how much mass is lost from the
secondary and deposit this material with the appropriate entropy at
the bottom of the stream. In this phase, the mass-transfer rate from
the secondary is still determined by the frictional angular-momentum
loss the immersed binary experiences, and reaches up to
100\,M$_\odot\,$yr$^{-1}$ in the final phase.

Since hydrogen is deposited in layers with temperatures $T > 10^8\,$K,
an extended nuclear reaction network, including all important weak
interactions, is included using the REACLIB library (Thielemann,
Truran \& Arnould 1986).  This library provides the data for reaction
rates which can be applied for a wide range of temperatures
($T=10^{\ 7}-10^{10}$\,K) for all elements up to $^{84}$Kr. For heavier
elements, the library has been updated for neutron-capture
cross-sections and half-lifes of beta-decays for the s-elements from
$^{85}$Rb up to $^{209}$Bi. The data for these cross-sections were
taken from the work of Beer, Voss and Winters (1992) and have been
interpolated using the method of minimal squares to fit the formulae
for reaction rates used in REACLIB. The data for the half lives of the 
beta-decays were taken from Newman (1978). We note that, in
explosive phases, the nuclear timescales of many reactions are shorter
than the convective turnover timescales in convective regions, and
thus the assumption of homogeneous mixing in the convective layers is
not valid. To treat the time-dependent nuclear burning in convective
zones, we therefore use a modified version of the two-stream formalism
developed by Cannon (1993).\footnote{Further details of the code can 
be found in Ivanova (2002) and in Ivanova \& Podsiadlowski (2010).}

\section{Explosive Merging and Common-Envelope Ejection}

\subsection*{\it The case of an 18\,M$_{\odot}$ primary}

To explore the conditions for explosive common-envelope ejection, we
performed a detailed numerical study of the merger of a 18\,$M_\odot$
primary with secondaries in the range of 1\,--\,5\,$M_\odot$. In all
cases, we assumed that the spiral-in phase started when the primary
had completed helium core burning and was ascending the red-supergiant
branch for the second time (so-called Case C mass transfer; 
Lauterborn 1970). Figure~2 shows the results of a typical calculation
which illustrates the evolution from the initial spiral-in phase to the point at
which the envelope is ejected.
%
To test the effects of convective overshooting, we used models with
0.25 pressure scale heights of convective overshooting and models
without convective overshooting, where the former produces a
H-exhausted core of 6.8\,$M_\odot$, 1.3\,$M_\odot$ larger than the
latter.
%
%
To examine the uncertainties due the modelling of the stream--core
interaction, we use two different parameters for the entropy
generation in equation~(28) of Ivanova et al.\ (2002), $k=0.2$ for
low-entropy generation and $k=0.4$ for high-entropy generation.
We note that, in the present study, we did not model how rotation
of the core affects the penetration depth (see section 5.3.2
of Ivanova et al.\ [2002]) and how this changes during the
stream--core interaction phase. We do not expect that this would
change our conclusions significantly.

\begin{figure}
\begin{center}
\includegraphics[width=0.47\textwidth]{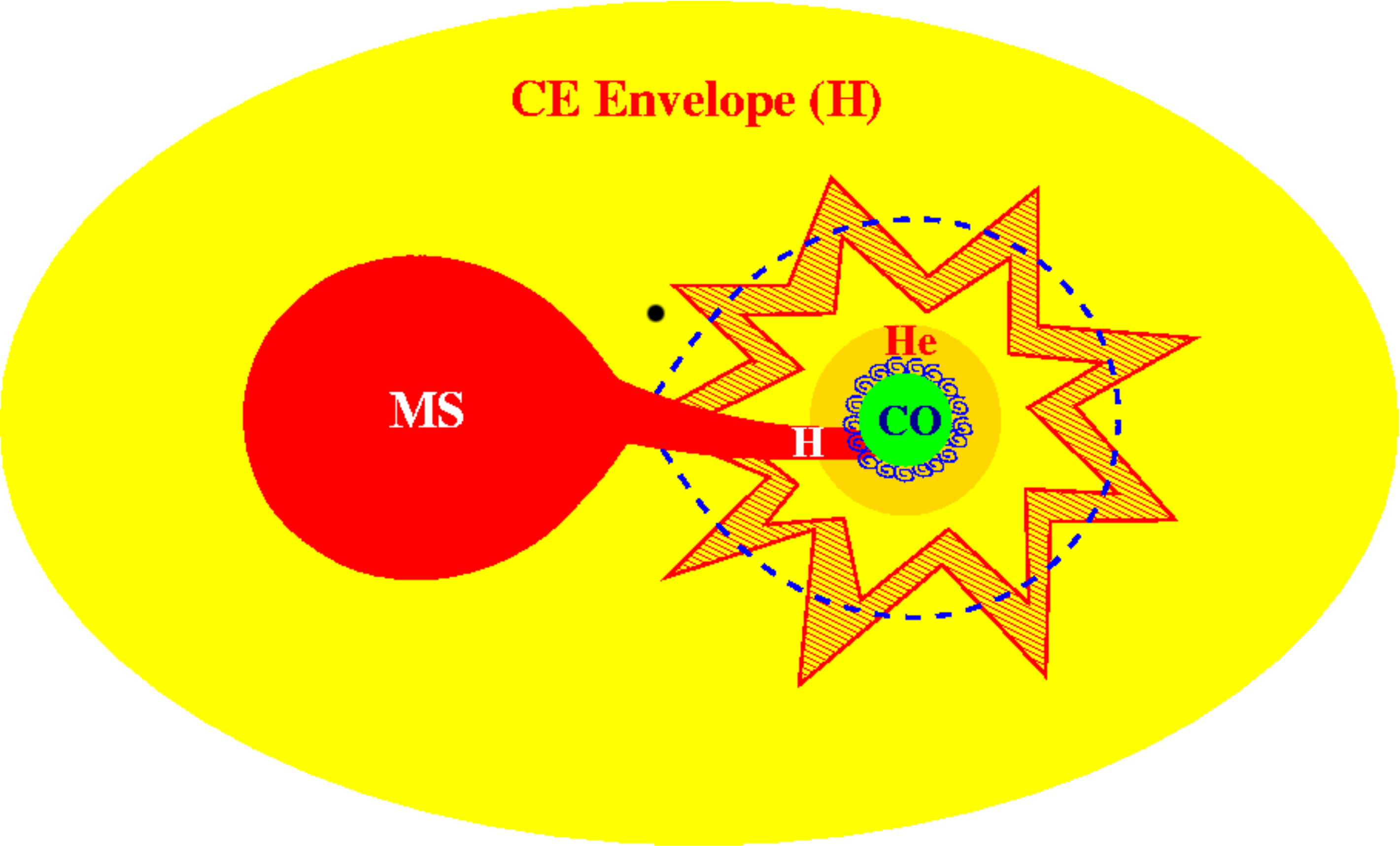}
\end{center}
\caption{Schematic illustration of the process of explosive
common-envelope ejection. The H-rich stream from the Roche-lobe-filling
immersed companion penetrates deep into the core of the primary, mixing
hydrogen into the helium-burning shell. This leads to a thermonuclear
runaway ejecting both the helium shell and the hydrogen-rich envelope
and leaving a bare CO core.}
\end{figure}

\subsection*{\it The initial set-up at the start of the merger}

At the end of the spiral-in, when the secondary starts to fill its
Roche lobe within the common envelope, the secondary itself is located
partially within the outer convective zone (OCZ) of the primary,
although the inner Lagrangian point, $L_1$, is within the radiative
zone. The layer of envelope material at the position of the secondary
has expanded because of the frictional energy input from the spiral-in
and has a density roughly two orders of magnitude less than before the
spiral-in, with a typical range from about $10^{-7}\,{\rm g\,cm^{-3}}$
to about $10^{-5} \,{\rm g\,cm^{-3}}$.  The spiral-in does not affect
the structure of the primary's core (the CO core plus the surrounding
He shell), it only affects the structure of the radiative hydrogen
zone between the OCZ and the He shell ($\Delta M_{\rm H,rad}$), since
the upper part of this zone has been heated and may have become
convective.  For example, in the case of a 5\,$M_\odot$ donor inside
an 18\,$M_\odot$ primary, the total mass of the OCZ is increased by
$0.4\Msun$, while, in the case of a $2\Msun$ donor, it is only
increased by $0.04\Msun$.  It should be noted that, in the first case,
the initial $\Delta M_{\rm H,rad}$ is 0.13\,$M_\odot$ for our standard
case with convective overshooting. Without convective overshooting, it
would be $0.7\,M_\odot$.  The corresponding radial sizes of this zone
($\Delta R_{\rm H,rad}$) are $1.6\times 10^{11}\,$cm and $2.5\times
10^{11}$\,cm for the case with and without convective overshooting,
respectively.

\subsection*{\it The stream behavior and the surrounding medium}

Once the stream leaves the secondary, the depth of its penetration
depends on the initial entropy of the stream material and on how much
this entropy is modified during the stream's penetration into the
primary's core by shocks (Ivanova et al.\ 2002). The latter is a
function of the density contrast between the stream and the
surrounding matter and depends on how strongly the medium itself is
stratified. A larger density gradient in the primary's material near the stream leads to
more  entropy generation in the stream. The stream will stop its
fall when the ambient pressure becomes comparable to the
stream pressure.

In the test cases, the penetration depth is two times smaller in the
model with large entropy generation (with $k=0.4$) than in the model
with $k=0.2$. The distance to the He shell from $L_1$ is also
larger. For a secondary that is more massive than $\sim 3\,M_\odot$,
the stream is unable to penetrate into the He shell since the
density contrast between the stream and the ambient medium is lower
and the stream entropy higher.

\subsection*{\it The response of the He shell to the injection of
hydrogen}

\begin{figure*}
\includegraphics[width=17.5cm]{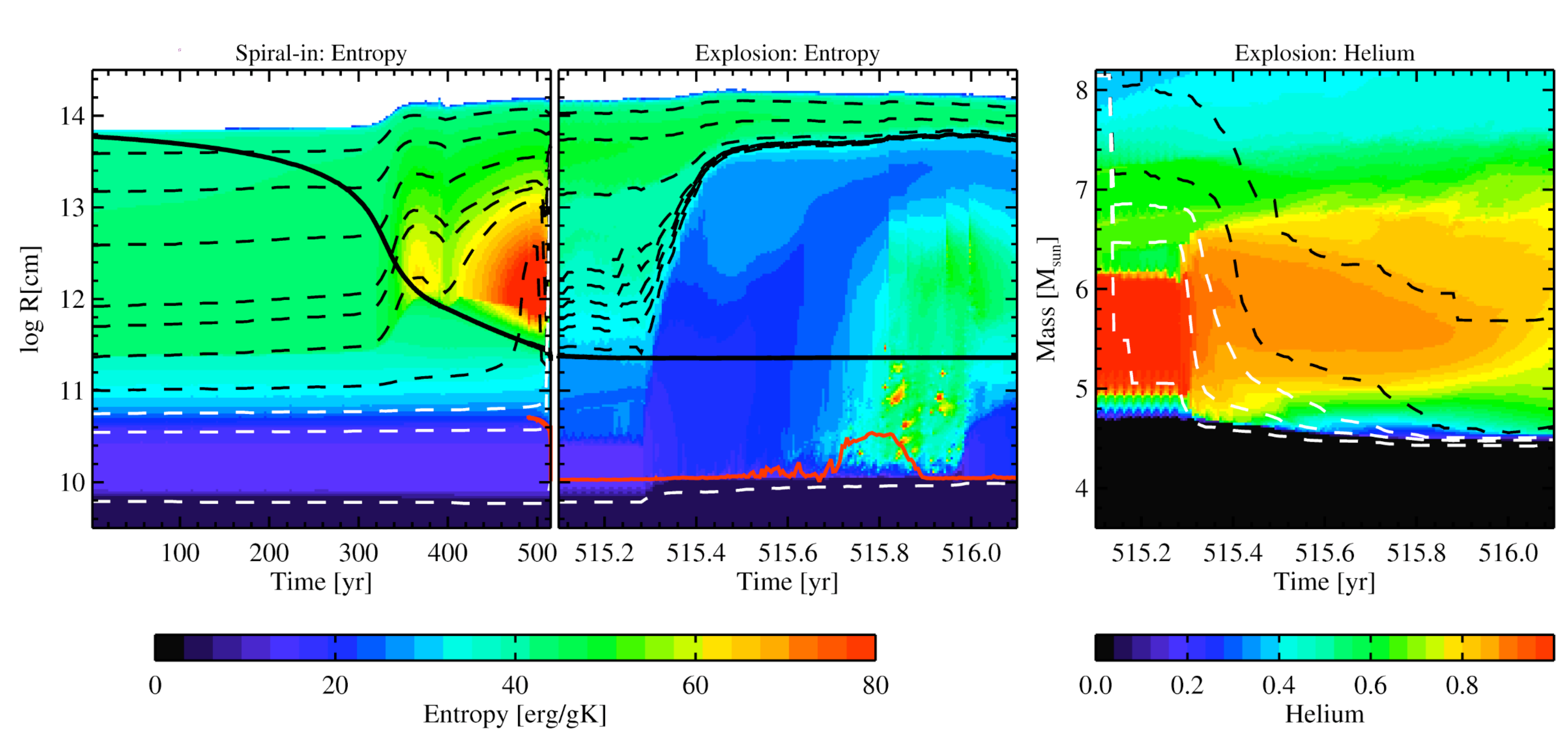}
\caption{Simulations of the spiral-in of a $2\,M_\odot$ star inside
  the envelope of a 18\,$M_\odot$ red supergiant (after He core
  burning) up to the phase where the H/He envelope is being
  ejected. The left and medium panels show the time-evolution of the
  specific entropy profile as a function of radius; {\it left panel:}
  from the start of the spiral-in up to the beginning of the He shell
  explosion phase; {\it middle panel:} during the final year around
  the He shell explosion.  The right panel shows the time-evolution of
  the helium profile as a function of mass during the last year.  The
  black solid curve in the left and medium panels shows the orbital
  position of the secondary, the solid red curves show the depth of
  the stream penetration.  The dashed curves are curves of constant
  mass and correspond to $M=14,9, 7.30, 6.96, 6.92, 6.88, 6.84, 6.80,
  6.72, 4.457\,M_\odot$ (from the top).  The dashed curves in the
  right panel are curves of constant radii and correspond to $\log
  R/{\rm cm} = 13.6, 13, 12,11, 10$ (from the top).  The stream
  penetration efficiency parameter for this model is $k=0.2$.}
\end{figure*}

In the layer where the hydrogen-rich stream penetrates into the
helium-dominated layer (see the middle and right panels of Figure~2),
the stream entropy is higher than that of the surrounding material. As
a consequence, both the entropy and the hydrogen mass fraction $X$
have negative gradients. This leads to the formation of a local
convective zone that rapidly expands outwards.  As the stream
continues to hit the core, it penetrates progressively deeper into the
core for several reasons: first, the entropy of the donor material
decreases as material from deeper inside the star is transferred.
Second, the outer layer of the He shell starts to expand, reducing the
density in the outer part of the He shell. Third, the mass transfer
rate steadily goes up, and, fourth, the $L_1$ point moves closer to
the helium core.  In the case where the initial stream does not
penetrate deeply enough into the He-rich layer and is unable to create
a negative entropy gradient in the ambient medium, a convective zone
still develops in the region where the stream is being dissolved, but
more slowly (the stream's entropy is still higher than the ambient
medium and stream energy is being deposited, further increasing the
entropy).  This convective zone propagates inwards as the stream
penetrates deeper.  We found that in some models this zone can connect
with the OCZ; however, it either never penetrates deeply enough to
connect to the convective helium-burning shell, or it starts efficient
steady hydrogen burning, leading to a drop of the temperature in the
He convective shell.

%
%

\subsection*{\it The conditions for explosive CE ejection}

Based on these simulations, it seems that the main condition for
experiencing explosive CE ejection (ECEE) is that the stream is able
to penetrate deeply enough into the He-rich zone from the very
beginning such that both a negative entropy and a negative composition
gradient are created.  This favours models with convective
overshooting and low-entropy generation. It also favours lower-mass
secondaries with lower entropies. For example, for $M_2=2\,M_\odot$,
after leaving $L_1$, the stream can penetrate as much as
2.5\,$R_\odot$ of the core, while for $M_2=5\,M_\odot$ it stops
further from the primary center, passing through only 1\,$R_\odot$.
An alternative criterion is that $\Delta M_{\rm H,rad}$ is small:
   $\Delta M_{\rm H,rad}\le 0.2\,M_\odot$.

Based on our test calculations, we conclude that the conditions for a
successful explosive common-envelope ejection are satisfied for
secondary masses up to 3\,$M_\odot$ in the models with convective
overshooting. A lower limit on the secondary mass is roughly given by
1\,$M_\odot$ where the energy released in the spiral-in does not
provide enough energy to lead to an expansion of the envelope
(i.e. lead to a ``slow'' merger); in that case the merger is expected
to occur on a dynamical timescale.

With these constraints to guide us, we find that the ECEE conditions
are found for primaries with masses up to 40\,$M_\odot$. If the
minimum mass for black-hole formation is 25\,$M_\odot$, this leads to
CO cores in the range of 6.5\,--\,13\,$M_\odot$ (using our standard
convective overshooting parameter).

The amount of mass transferred from the secondary before the CE
ejection also depends on the secondary's mass: it is about
1\,M$_\odot$ for a 3\,$M_\odot$ secondary, 0.8\,$M_\odot$ for a
2\,$M_\odot$ secondary and 0.2\,--\,0.3\,$M_\odot$ for a 1\,$M_\odot$
secondary (but this depends somewhat on the stream entropy parameter).

\subsection*{\it The final stage}

While the initial spiral-in and the early merger phase take place on a
timescale of $\sim 100\,$yr, once the H-rich stream connects to the
helium-core-burning shell, the evolution accelerates.  The overall
duration of the explosive phase is $\sim 1/4$\,yr, although 90\% of the
energy is released in the last few days. At the end of our
calculations, the outer parts of the core expand with a velocity
exceeding the local escape velocity, and typically
0.03\,--\,0.06\,$M_\odot$ of stream material has been burned
explosively. The nuclear energy that has been released ($\ga 2
\times 10^{50}$ ergs) exceeds the binding energy of both the He-rich
layer and the common envelope by about a factor of 2.

By the time the envelope is ejected, we find that the CO core has been
moderately spun-up by the accretion of angular momentum and transport
into the core during the pre-explosive core-accretion phase. The
characteristic specific angular momentum of the core is $\sim
10^{16}\,$cm$^2\,$s$^{-1}$, at the low end of what is required in the
collapsar model for LGRBs.  However, further spin-up is possible, since
even after the ejection of the envelope, mass transfer is likely to
continue and the system is close enough for tidal spin-up to operate.\footnote{
Detmers et al.\ (2008) found that, in a sufficiently close He-star
binary (with $P_{\rm orb}\la 10\,$h), the core can be spun up
significantly on a timescale of $\sim 10^{4}$\,yr (also see Brown et
al.\ 2007). However, unlike the case considered by Detmers et
al.\ (2008), the remaining lifetime of the primary after case C mass
transfer is short enough that wind mass loss will not cause
significant widening of the system which would then tidally spin down
the star again.}

Indeed, despite the ejection of the helium shell, the system does not
widen appreciably, if at all, because there is still enough frictional
angular-momentum loss to keep the secondary in Roche-lobe
contact. This suggests that the system will continue to transfer mass
after the CE ejection.\footnote{We note, however, that there are enough
  uncertainties in our calculations that we cannot rule out that the
  secondary will be out of Roche-lobe contact after the CE
  ejection. This will ultimately require full three-dimensional
  stellar-structure calculations.}  This mass transfer occurs on a
thermal timescale since the secondary is highly out of thermal
equilibrium, in fact has a radius smaller than its equilibrium radius,
and will try to expand towards a new thermal-equilibrium stage.

The remaining evolution of the primary's CO core to core collapse is
expected to be of order a few $10^3$ to at most $\, 10^4\,$yr (in the
18\,$M_\odot$ simulation it was 3000\,yr). We note that this time is
significantly longer than it would have taken the core to core
collapse without the ejection of the He layer because of the
signifiant core expansion and cooling.

\section{Application to LGRBs and short-period black-hole binaries}

In order to assess the importance of ECEE events for LGRBs and
low-mass black-hole binaries, we first estimate in this section
the expected rate of these events and then present detailed
binary population synthesis results that simulate the properties
of the resulting black-hole (BH) binaries.

\begin{figure}
\begin{centering}
\includegraphics[width=8.4cm]{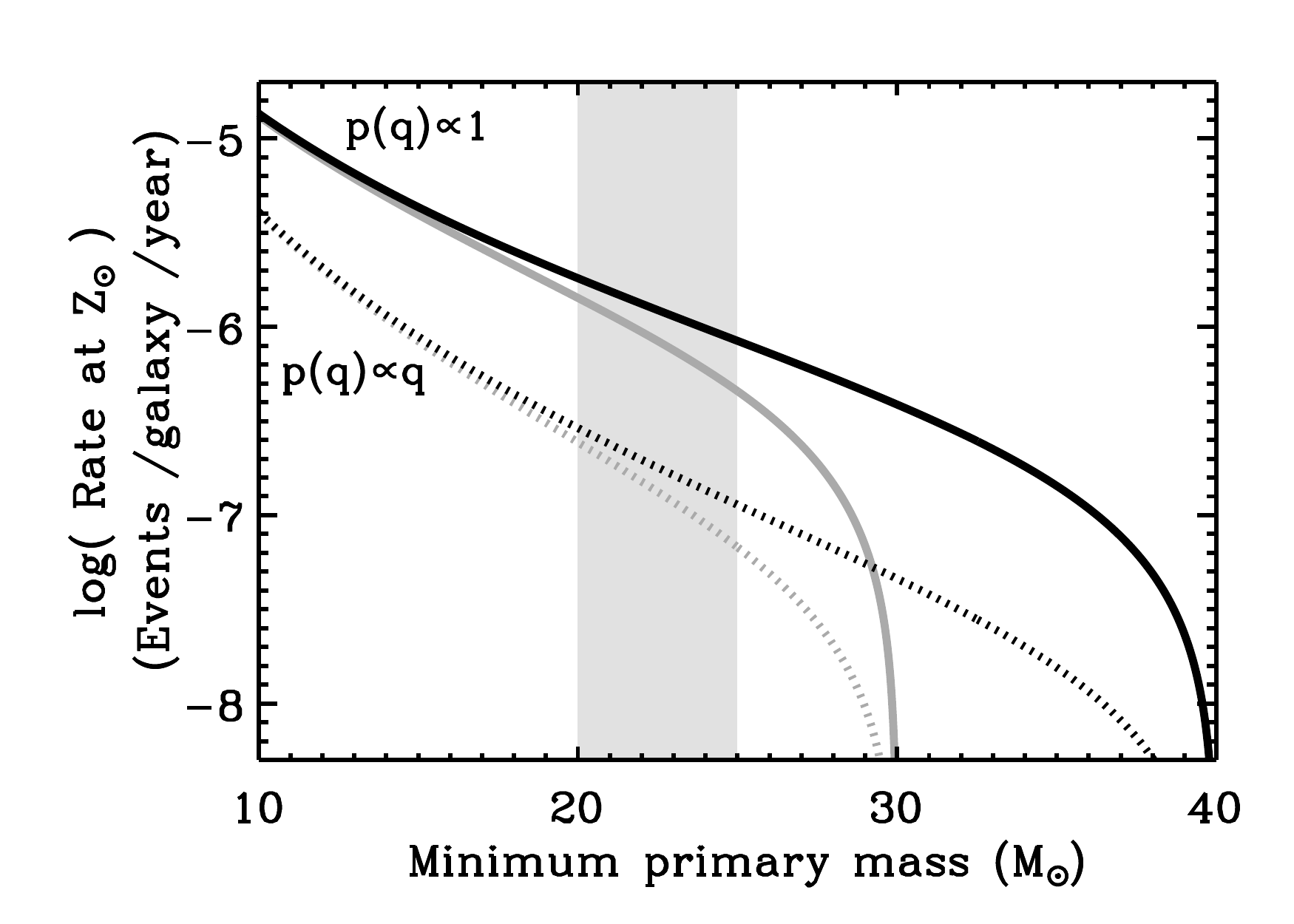}
\caption{
\label{fig:IMFtimesIMRD}
The cumulative rate for ECEE events as a function of the minimum
primary mass required. The black curves assume a maximum primary mass
of 40\,$M_{\odot}$ whilst the grey curves take a maximum primary mass
of 30\,$M_{\odot}$. The solid curves assume a flat initial mass-ratio
distribution ($p(q) = {\rm constant}$), the dotted curves use $p(q)
\propto q$.  The light grey shading indicates a plausible range for
the minimum mass of a single star which can form a black hole at solar
metallicity.  For these estimates we assumed that 1\,\% of binaries
have an orbital period that leads to Case C mass transfer. The actual
rate could be higher by an order of magnitude at low metallicity, and
the minimum mass for BH formation could be lower.  }
\end{centering}
\end{figure}

\subsection{The ECEE Rate}

Using the results discussed in the previous section, it is
straightforward to estimate the fraction of binaries that experience
explosive CE ejection. This depends mainly on the secondary mass
and the condition that the systems experience Case C mass transfer.

Figure \ref{fig:IMFtimesIMRD} shows the cumulative rate for ECEE
events as a function of the minimum primary mass for different
assumptions about the maximum mass of the primary and different
mass-ratio distributions. Here, we assume that 1\,\% of all binaries
experience Case C mass transfer at solar metallicity. The primary
masses $M_{\rm 1}$ are chosen from an initial mass function $p(M_{\rm
  1}) \propto M_{\rm 1}^{-2.7}$ (Kroupa, Tout \& Gilmore 1993). The
rates are nomalised to a core-collapse SN rate of 1 per century (i.e.,
assume that one star more massive than $9\,M_{\odot}$ is formed per
century in a typical galaxy).  This probably underestimates the
core-collapse supernova rate in the Milky Way (Cappellaro, Evans \&
Turatto 1999).  We also assumed that all massive stars form in
binaries (Kobulnicky \& Fryer 2007), with a distribution of initial
separations $a$ that is flat in $\log a$ between 3 and
$10^{4}~R_{\odot}$.

A significant source of uncertainty in the ECEE rates is the initial
mass-ratio distribution $p(q)$, where $q = M_{\rm 2}/M_{\rm 1}$ is the
mass ratio and $M_{\rm 2}$ is the mass of the secondary. In Figure
\ref{fig:IMFtimesIMRD} we consider a distribution where $p(q) = {\rm
  constant}$ and one where $p(q) \propto q$. 


In using the ECEE mechanism to produce short-period black-hole X-ray
binaries, we require the primary to form a BH. The light grey shading
in Figure~\ref{fig:IMFtimesIMRD} indicates the expected minimum mass
for black-hole formation, which is generally believed to be in the
range of 20\,--\,25\,$M_\odot$ (see, e.g., Fryer \& Kalogera 2001 and
further references in PRH). Black-hole formation is also generally
assumed to be a requirement for the formation of LGRBs, though we note
that rapid rotation at core-collapse for lower mass cores may lead to
the formation of magnetars (e.g., Thompson \& Duncan 1993; Akiyama et
al. 2003). Magnetar formation has also been proposed as a channel for
LGRBs (e.g., Wheeler, H\"oflich \& Wang 2000; Burrows et al.\ 2007;
Uzdensky \& MacFadyen 2007; Bucciantini et al.\ 2008), which would
increase the potential LGRB event rate from this proposed mechanism.
Because of the steepness of the primary mass function, the upper limit
of 40\,$M_{\odot}$ indicated by our hydrodynamic calculations is of
lesser importance to the rates.

The fraction of binary systems that experience Case C mass transfer is
strongly dependent on the metallicity and the assumptions about the
wind-loss rate from the primary star (see Justham \& Podsiadlowski
2010), but for Z=0.02, 1\,\% is a reasonable estimate. On the other
hand, if the Nieuwenhuijzen \& de Jager (1990) wind-loss rates are
correct, only primary stars with masses $ \la 30\,M_{\odot}$ can
experience Case C mass transfer. This mass goes up to
$\simeq 40\,M_\odot$ if the Nieuwenhuijzen \& de Jager rate is
reduced by a factor $1/3$, to take into account that the the empirical 
rates are not corrected for the effects of wind
clumping (e.g., Moffat \& Carmelle 1994; Fullerton, Massa \& Prinja
2006).  However, as Figue~3 shows, the uncertainty in this upper mass
limit is not a significant contribution to the uncertainty in the
ECEE rate.

Hence, if we only allow primary masses of $25\,{M_{\odot}} \leq
M_{\rm 1} \leq 40\,{M_{\odot}}$ to produce LGRBs through the ECEE
channel, and take $p(q) = \mbox{constant}$, we obtain a LGRB rate of
$\sim 10^{-6}$ gal$^{-1}$ yr$^{-1}$. This is consistent, though at the
lower end, of recent LGRB rate estimates (e.g.,  Podsiadlowski et
al.\ 2004). There is some evidence that LGRBs prefer lower metallicity
(e.g., Fruchter et al.\ 2006; Wolf \& Podsiadlowski 2007; Modjaz et
al.\ 2008). This is consistent with the ECEE scenario, since the rate
of case C mass transfer increases with decreasing metallicity, by up
to at least a factor of 10 at extremely low metallicity ($Z\la
0.0001$; Justham \& Podsiadlowski 2010). Given the uncertainties in
these estimates, we conclude that the ECEE channel could account for
at least a significant subset of the LGRB rate and possibly all of it.

\subsubsection*{Consistency between the LGRB rate and the
short-period BH binary birthrate}

ECEE links LGRBs with short-period black-hole X-ray binaries.  Wijers
(1996) and Romani (1998) estimated the Galactic population of
short-period BH XRBs to be at least $\sim 1000$. For a birthrate of
$\sim 10^{-6}$ gal$^{-1}$ yr$^{-1}$, this requires a plausible mean
system lifetime of $\sim 1\,$Gyr. This assumes that the majority of
systems will not be disrupted by the supernova explosion and will
reach Roche-lobe contact in a reasonable time. In the following
subsection we simulate the properties of the predicted X-ray binary
population from this channel.

\subsection{Method \& Assumptions}

For the population synthesis calculation, we randomly selected a set
of $5 \times 10^{\rm 5}$ binaries with initial primary masses $M_{\rm
1}$ between 25 and 40\,$ M_{\odot}$ and initial secondary masses
$M_{2}$ between 1 and 3\,$ M_{\odot}$. We again adopted an initial
mass function $p(M_{\rm 1}) \propto M_{\rm 1}^{-2.7}$ and a flat mass-ratio
distribution. After the CE ejection, the secondary is assumed to
be filling its Roche lobe, although in reality it may be slightly
underfilling it because of the mass loss associated with the CE ejection.

The core mass of the primary $M_{\rm core}$ can be related to the initial
primary mass according to the simple estimate
\begin{equation}
(M_{\rm core} / {M_{\rm \odot}}) = 0.12~\left(M_{\rm 1}/ 
{ M_{\rm \odot}}\right)^{1.35}.
\end{equation}
(Hurley et al.\ 2000).
This yields post-ECEE remnant masses in the range of
9.3\,--\,17.5\,$M_\odot$ for initial primary masses between 25 and
40\,$M_\odot$.  The secondary, having lost mass during the CE phase,
leaves a donor star of mass $M_{\rm don}$, where
\begin{equation}
(M_{\rm don}/ { M_{\rm \odot}}) = 0.2+0.6\,(M_{2}/ { M_{\rm \odot}})
\end{equation}
(in accordance with the results in Sect.~3).
The orbital period and component masses are assumed to be unchanged
until the core explodes as a supernova. After the supernova, the core
becomes a BH; we determine the mass of the BH using the results of
Fryer \& Kalogera (2001), as formulated in Belczynski et
al. (2008).\footnote{Specifically, their equations 1, 2 and 4.}

Mass loss from the system and any kick imparted to the black hole
during the supernova will change the orbital period and eccentricity
(see, e.g., Brandt \& Podsiadlowksi 1995; Kalogera 1996).  We assume
that the supernova kicks are randomly oriented, and sample the kick
distributions assuming that it can be described by 
a Maxwellian distribution with a velocity dispersion $\sigma_{\rm
  k}$ (for a discussion of supernova kicks in the formation of black
holes, see Brandt, Podsiadlowski \& Sigurdsson 1995; PRH). We explored
a range of values for $\sigma_{\rm k}$, including zero kick velocity
and kick velocities with the same kick {\em momentum} as found for 
single neutron stars (Hobbs et al.\ 2005).

If the system is not disrupted in the supernova, we assume that it
circularises whilst conserving angular momentum. We also allow
the system to survive even if the Roche-lobe radius of the donor star
is as small as 90\% of the ZAMS radius for the post-ECEE donor mass to
take into account the fact that the secondary is likely to be still
undersized and expanding at this stage, as indicated by our ECEE
calculations. We assume that a short period of mass transfer driven by
this expansion will increase the orbital period of the system until
the star is in equilibrium, and assume, for simplicity, that the mass
lost by the donor during that phase is negligible. The donor star is
now assumed to be still essentially unevolved.

We weight the output of this algorithm by the initial mass function
and initial mass-ratio distribution. This produces a probability
distribution for the formation of post-supernova black-hole
binaries. In order to simulate the population of black-hole X-ray
binaries, we combine these formation probabilities with a grid of
binary evolution sequences.


\subsubsection*{Binary evolution grid}

To follow the X-ray binary phase, we calculated a grid of 936 binary
evolution sequences using an updated version of Eggleton's stellar
evolution code (e.g., Eggleton 1971; Pols et al.\ 1995) and a
metallicity of 0.02. This grid was equally spaced in 13 donor masses
(0.8 to 2.0\,$ M_{\odot}$), 13 accretor masses (5 to 17\,$M_{\odot}$), 
and 12 values for the initial period. The first model is
placed into contact at the start of its evolution, with an orbital
period $P_{\rm orb}=P_{\rm ZAMS}$. The second begins in a binary with
$P_{\rm orb}/P_{\rm ZAMS}$=1.1, and the remaining ten orbital periods
are equally spaced in $\log\, (P_{\rm orb}/P_{\rm ZAMS})$ between 0.1 and
1.0 (where $P_{\rm ZAMS}$ is the orbital period at which the donor is
filling its Roche lobe when unevolved).\footnote{Note that $P_{\rm
    ZAMS}$ is a function of donor mass, accretor mass and donor
  radius. Due to angular-momentum loss, the models with initial $P_{\rm
    orb}>P_{\rm ZAMS}$ have shorter periods by the time they reach
  contact.}

The evolution of the systems is driven by angular-momentum loss. We
include losses due to both gravitational wave radiation and magnetic
braking, adopting the formalism of Ivanova \& Taam (2003) for
secondaries with $M_2 < 1.5\,M_\odot$ (see also Ivanova \& Kalogera
2006). We consider only systems which reach contact before the end of
the main sequence and have a maximum age for each system of 13 Gyr; we
varied this maximum age and found it to have a relatively minor
effect.

The formation probabilities produced by our population synthesis are
all distributed among the eight members of the three-dimensional grid
nearest to them, i.e. to a cube of gridpoints around each system,
using trilinear interpolation to determine the weights attributed to
each grid point. Each sequence in our binary evolution grid then
posesses a formation likelihood, and the grid sequences are binned and
combined in proportion to these formation probabilities. Within each
sequence, the statistical weight of each evolutionary timestep is
proportional to the duration of that timestep. This leads to a
`steady-state' X-ray binary population, assumed to be representative
of the distribution in the Milky Way at the current epoch.



\begin{figure}
\begin{centering}
\includegraphics[width=8.4cm]{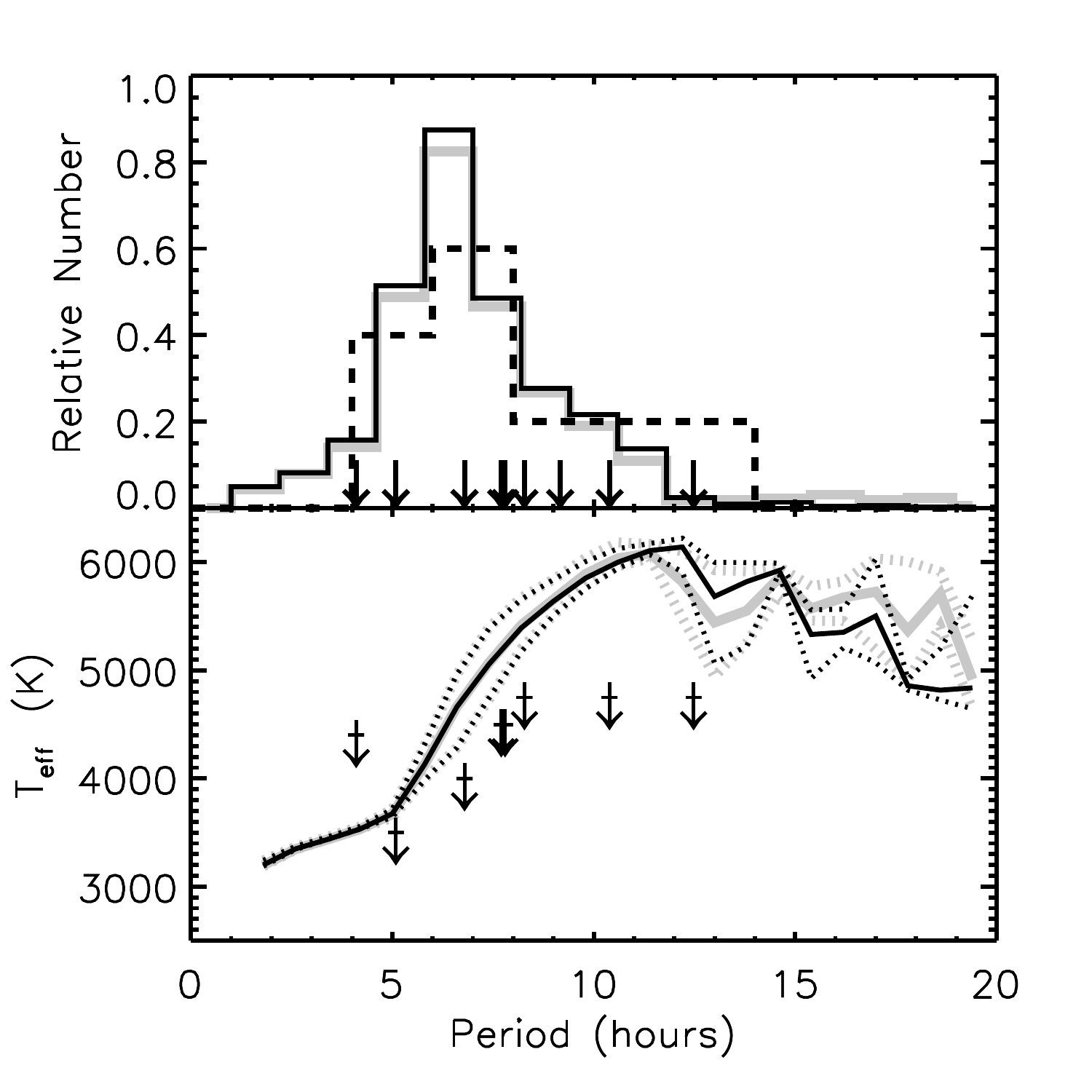}
\caption{
 \label{fig:TeffPeriodDists} Orbital period distribution (upper panel)
and distribution of effective temperatures of the secondaries for two
population synthesis simulations. The thick grey curves assume a
supernova kick dispersion of $\rm \sigma_{\rm k}=200~km~s^{-1}$, and
the solid black curves correspond to no kick. Both simulations assume a flat
initial mass ratio distribution ($p(q) \propto 1$) (changing this to
$p(q) \propto q$ makes little difference to the shape of the
distribution, but changes the overall normalisation).  In the top
panel, the dashed curve represents a histogram of the orbital periods
of known systems (also marked with arrows). The normalisation of the
histogram of observed systems is arbitrary.  The shapes of the period
distributions are reasonably well matched. The lower panel shows the
median effective temperature (solid curves) and upper and lower
quartiles (broken curves) of the populations as a function of orbital
period. We also mark the maximum effective temperatures consistent
with the spectral types of the known systems (crosses with arrows).}
\end{centering}
\end{figure}

\subsection{Results}

Figure \ref{fig:TeffPeriodDists} shows a selection of our population
synthesis calculations, which reproduces the shape of the observed
orbital-period distribution fairly well, for $\sigma_{\rm k}={\rm 200\,km
\,s^{-1}}$ and for no kick. However, the donor temperatures do not
easily match the observations, unless either our model effective
temperatures or our conversion from spectral type to effective
temperatures is incorrect.

Note that these donor temperatures give a significantly better match
to the observations than systems which initially contained
intermediate-mass donors would provide (see, e.g., Justham et
al.\ 2006), and that this apparent mismatch in effective temperature
seems to be a problem for all models, unless the donor stars in
short-period BH binaries are pre-main sequence stars (Ivanova
2006). Hence this problem with effective temperatures appears to be a
generic problem for models of these systems.

\section{Conclusions}

In this paper, we have presented a new mechanism for the ejection of a
common envelope where the energy source is not orbital energy but
nuclear energy. This provides a new channel to produce plausible
progenitors of short-period black-hole binaries and long-duration
GRBs.

We have also demonstrated that this scenario may be able to explain
both the origin and the main properties of short-period black-hole
binaries, such as the period distribution, although the distribution of
spectral types is still not fully satisfactory.

Two of the most attractive features of the explosive CE ejection
scenario are that (1) it leads to the ejection of both the hydrogen
and the helium layer, explaining why all LGRB supernovae to-date have
been classified as Type Ic supernovae, and (2) this ejection occurs
late in the evolution of the star; hence the progenitor will only
experience a short Wolf-Rayet phase, in which it will not be spun down
significantly by wind mass loss.  This may also explain why extended
Wolf-Rayet wind bubbles are not being found around LGRBs; e.g., in the
case of GRB 021004, van Marle, Langer \& Garcia-Segura (2005) found
that the Wolf-Rayet phase had to last less than $10^4$\,yr; this is
consistent with the explosive CE-ejection scenario, since it always
requires late case C mass transfer. This also predicts that such
Wolf-Rayet bubbles are terminated by a dense shell from the ejected
common envelope.  Our rate estimate for this channel ($\sim
10^{-6}\,$yr$^{-1}$) implies that this can produce a significant
fraction of all LGRBs; this rate should be higher at lower
metallicity, because case C mass transfer is expected to be more
common at lower metallicity.  Unlike some of the single-star
progenitor models for LGRBs (Yoon \& Langer 2005; Woosley \& Heger
2006; Yoon, Langer \& Norman 2006), LGRBs may occur even at solar
metallicity, but they are expected to be more common at low
metallicity. Indeed, there is some evidence now that LGRBs can also
occur in super-solar host galaxies (see, e.g., Levesque et
al.\ 2010). The different metallicity biases may provide a possible
way of distinguishing between these two different scenarios.

Explosive CE ejection also operates for lower-mass primaries that are
expected to produce neutron stars rather than black holes.  It is
tempting to associate these with rapidly rotating neutron stars and
possibly magnetars, for which SN 2006aj, which was associated with the
X-ray flash GRB 060218, may provide an observed example in nature
(Mazzali et al.\ 2006).

\bigskip

\noindent{\bf ACKNOWLEDGEMENTS}
\medskip

\noindent
NI gratefully acknowledges support from the NSERC of
Canada and from the Canada Research Chairs Program.  SJ is partially
supported by the National Science Foundation of China under grant
numbers 10903001 and 10950110322, and by the Chinese Postdoc Fund
(award number 20090450005).

\end{document}